# Magneto-thermopower in the Weak Ferromagnetic Oxide CaRu$_{0.8}$Sc$_{0.2}$O$_3$: An Experimental Test for the Kelvin Formula in a Magnetic Material

Takafumi D. Yamamoto[1]*, Hiroki Taniguchi[1], Yukio Yasui[2], Satoshi Iguchi[3], Takahiko Sasaki[3], and Ichiro Terasaki[3]

[1] *Department of Physics, Nagoya University, Nagoya 464-8602, Japan*
[2] *Department of Physics, Meiji University, Kawasaki 214-8571, Japan*
[3] *Institute for Materials Research, Tohoku University, Sendai 980-8577, Japan*

We have measured the resistivity, the thermopower, and the specific heat of the weak ferromagnetic oxide CaRu$_{0.8}$Sc$_{0.2}$O$_3$ in external magnetic fields up to 140 kOe below 80 K. We have observed that the thermopower $Q$ is significantly suppressed by magnetic fields at around the ferromagnetic transition temperature of 30 K, and have further found that the magneto-thermopower $\Delta Q\,(H, T) = Q(H, T) - Q(0, T)$ is roughly proportional to the magneto-entropy $\Delta S\,(H, T) = S(H, T) - S(0, T)$. We discuss this relationship between the two quantities in terms of the Kelvin formula, and find that the observed $\Delta Q$ is quantitatively consistent with the values expected from the Kelvin formula, a possible physical meaning of which is discussed.

## 1. Introduction

Layered cobalt oxides have recently attracted much attention because they exhibit an unusually large thermopower with relatively low resistivity,[1-3] which is difficult to explain from simple band theories. On the basis of a localized electron picture, the spin and orbital degrees of freedom due to the strong electron correlation remain in these materials, being considered as a source of the enhanced thermopower.[4-8] In the related cobalt oxide Sr$_3$YCo$_4$O$_{10.5}$,[9] the thermopower changes with the spin state of the Co$^{3+}$ ions. Since the Co$^{3+}$ ions do not contribute to the electrical conduction, the thermopower can detect the background magnetic entropy in magnetic oxides, and give information complementary to the conductivity.

To gain a deeper insight into the relationship of the thermopower to the magnetism in magnetic materials, we need a suitable candidate in non-cobalt oxides. For this purpose we focus on the orthorhombic perovskite ruthenate CaRu$_{0.8}$Sc$_{0.2}$O$_3$, which shows a weak ferromagnetic glassy state with metallic conduction.[10, 11] Figure 1(a) shows the inverse magnetic susceptibility $H/M$ of CaRu$_{1-x}$Sc$_x$O$_3$.[10] Whereas CaRuO$_3$ ($x = 0$) is paramagnetic down to the lowest temperature, CaRu$_{0.8}$Sc$_{0.2}$O$_3$ ($x$ = 0.2) shows a steep drop in $H/M$ at around $T_c$ ~ 30 K, indicating the onset of the ferromagnetism. The Weiss temperature for $x$ = 0.2 is almost zero [see the solid line in Fig. 1(a)], which strongly indicates that a mean-field picture is no longer valid. As shown in Fig. 1(b), the resistivity $\rho$ for $x$ = 0.2 is non-metallic ($d\rho/dT$ < 0). The non-metallic behavior is, however, modest because $\rho$ increases only by a factor of 30 from 300 down to 4.2 K. Such a mild temperature variation of $\rho$ is a hallmark of disordered metals.[12-14] We have proposed a non-uniform magnetic state in which the Ru$^{5+}$ ions induced by the doped Sc$^{3+}$ ions dominate ferromagnetism, and the Ru$^{4+}$ ions are responsible for electrical conduction.

Here we report the magnetic field effects on the transport and thermodynamic properties of CaRu$_{0.8}$Sc$_{0.2}$O$_3$. We have found that the thermopower $Q$ is markedly suppressed by magnetic fields at around 30 K. This clearly suggests a substantial contribution of the magnetism to the thermopower. We further find that the magneto-thermopower $Q(H, T) - Q(0, T)$ is roughly proportional to the magneto-entropy $S(H, T) - S(0, T)$, and that the magneto-thermopower is consistent with the values expected from the Kelvin formula.[15] This is the first and unambiguous experimental evidence that the Kelvin formula is valid for the magnetic part of the thermopower of a magnetic material.

## 2. Experimental Details

A polycrystalline specimen of CaRu$_{0.8}$Sc$_{0.2}$O$_3$ was synthesized by a solid-state reaction method. A stoichiometric mixture of CaCO$_3$, RuO$_2$, and Sc$_2$O$_3$ of 99.9% purity was calcined in air at 900 ºC for 12 h. The calcined powder was then pressed into pellets after regrinding, and sintered in air at 1250 ºC for 48 h. Powder X-ray diffraction showed the orthorhombic structure with space group *Pnma* without any detectable impurity phases, consistent with our previous studies.[10]

Resistivity and thermopower were measured by a standard four-probe method and a steady-state two-probe technique, respectively. The data were collected between 2 and 90 K for various magnetic fields ($H$) in a Quantum Design physical property measurement system (up to 70 kOe) and a superconducting magnet (up to 150 kOe) at the High Field Laboratory for Superconducting Materials, Institute for Materials Research, Tohoku University. The magnetic field was applied perpendicular to the electrical current ($I$) and thermal gradient ($-\nabla T$). The specific heat ($C$) was measured by a thermal relaxation method from 0 to 90 kOe from 2 to 80 K using a Quantum Design PPMS equipped with a helium recycling system.

## 3. Results and Discussion

Figure 2 depicts the field dependence of the magnetoresistance MR($H$, $T$) = $\rho(H, T)/\rho(0, T) - 1$ at various temperatures. MR is negative below 40 K, and the magnitude increases with decreasing temperature, taking a maximum value of about −15% at 2 K in 150 kOe. In addition, MR($H$, 2 K) shows a hysteretic curve with a sign change in low magnetic fields. These features correspond to the magnetization $M$.[11] As shown in the inset of Fig. 2, the magnetic hysteresis curve strongly correlates with the MR($H$, 2 K) curve. When $M$ is equal to zero at a coercive field of 10 kOe, MR($H$, 2 K) takes the largest positive value.

A similar correlation between MR($H$, $T$) and $M$ is found in other ferromagnetic systems.[16, 17] This is explained in terms of a spin-dependent tunneling effect, where spin-polarized carriers undergo spin-dependent scattering at the boundaries between domains with different magnetization directions. In this tunneling process, the scattering rate depends on the relative angle of the magnetization directions of the adjacent domains, and hence changes with magnetic fields through magnetic domain rotation. Accordingly, the observed magnetoresistance probes the change in the scattering rate at the domain boundaries.

Figure 3 shows the temperature dependence of the thermopower of CaRu$_{0.8}$Sc$_{0.2}$O$_3$ in several magnetic fields. We find the negative magneto-thermoelectric effect below 80 K, where the thermopower is suppressed with increasing magnetic field. Unlike the resistivity, the magnetic field variation of the thermopower is most significant at around $T_c \sim$ 30 K. As shown in the inset of Fig. 3, MT($H$, $T$) = $Q(H, T)/Q(0, T) - 1$ takes a large value of about −15% at 30 K in 140 kOe. Note that the magnetoresistance is about −5% in the same temperature and magnetic field. Besides, no magnetic hysteresis is observed in MT($H$, $T$) at any temperature (not shown). MT($H$, $T$) has been often related to MR($H$, $T$),[18, 19] but this is not the present case, for MT($H$, $T$) is totally different from MR($H$, $T$).

As the thermopower is equal to the transport entropy (i.e., entropy per charge), we will compare the magneto-thermopower with the magneto-entropy. Figure 4(a) shows the specific heat divided by temperature, $C/T$, plotted against temperature for different magnetic fields. No jump at $T_c$ is seen in zero magnetic field, indicating a magnetic glassiness.[20-22] Our dynamic magnetic measurements[11] have suggested that the Sc-induced ferromagnetism results from a cluster glass state with a long relaxation time. A similar featureless specific heat has been reported in CaRu$_{1-x}$Ti$_x$O$_3$.[23]

One can barely see a tiny change in the temperature dependence around $T_c$ for 90 kOe: $C/T$ bends downwards (upwards) below (above) $T_c$. To make this clear, we show the relative change in $C/T$, defined as $\Delta C(H, T)/T = [C(H, T) - C(0, T)]/T$, in Fig. 4(b). We find that

$\Delta C(H, T)/T$ changes its sign at around $T_c$ and evolves systematically with increasing magnetic field. These are typical features of the magnetocaloric effect for conventional ferromagnets, corresponding to the suppression of the ferromagnetic fluctuation by external fields.[24, 25]

Here let us find the relationship of the magneto-thermopower with the magneto-entropy in $CaRu_{0.8}Sc_{0.2}O_3$. Figure 5(a) shows the magneto-entropy $\Delta S(H, T)/T = S(H, T) - S(0, T)$ calculated from $\Delta C(H, T)/T$ by using the following relation as

$$\Delta S(H,T) = \int_0^T \frac{\Delta C(H,T)}{T} dT. \quad (1)$$

A broad dip is seen around $T_c$, and becomes pronounced with increasing $H$. In high magnetic fields, $\Delta S(H, T)$ is visible far above $T_c$, because external fields suppress the ferromagnetic fluctuation. We show the magneto-thermopower $eQ(H, T) = e[Q(H, T) - Q(0, T)]$ in Fig. 5(b), where $e$ is the element charge. Remarkably, $eQ(H, T)$ resembles $\Delta S(H, T)$, implying a strong link between the two quantities.

We discuss this relationship on the basis of the Kelvin formula, which is proposed as a good approximate expression of the thermopower for various materials including correlated electron materials.[15] The Kelvin formula is given by

$$Q_K = -\frac{1}{e}\left(\frac{\partial \mu}{\partial T}\right)_{N,V}, \quad (2)$$

where $\mu$ is the chemical potential, $N$ is the particle number, and $V$ is the volume. This formula has the advantage that it describes the temperature dependence of the thermopower without restriction on the temperature range.[26, 27] As pointed out in several theoretical works,[28-30] this formula is more appropriate for incoherent charge transport. Thus, the title compound is a suitable candidate to test the validity of this formula because the non-metallic resistivity is a clear sign of incoherent transport.

The Gibbs-Duhem equation gives

$$-\left(\frac{\partial \mu}{\partial T}\right)_{N,P} = \frac{S}{N}, \quad (3)$$

where $P$ is the pressure. In solids far below the melting point, the constant-volume condition can be identified with the constant-pressure condition because the thermal expansion coefficient is sufficiently small. Then we arrive at $eQ_K = S/N$, and deduce that the magneto-thermopower and the magneto-entropy have the same relationship as

$$e\Delta Q_K = \frac{\Delta S}{N}. \quad (4)$$

Here we take $k_B$ per formula unit as the unit of $\Delta S$, and $N$ equals the number of carriers per formula unit. Figure 6 shows the correlation between $\Delta S$ and $\Delta Q$ below $T_c$, where all the data fall into a single curve within experimental uncertainties. The dotted line indicates $N = \Delta S/e\Delta Q = 1$, which means that the carrier number is nearly the same as the number of Ru ions, and is consistent with the measured Hall coefficient of $CaRuO_3$.[31]

We notice that the data gradually deviate from the line as the temperature increases. One possibility is that the thermal fluctuations disturb the magnetic field effects. The applied magnetic field in this study is an order of magnitude corresponding to the thermal energy of 10 K. Thus, it is more difficult to measure accurately the magnetic field variation at high temperatures. Another possible explanation is that the coupling between the carriers and the ferromagnetism grows with decreasing temperature. The relationship between $\Delta Q$ and $\Delta S$ would originate from a kind of magnon-drag effect, which is marked in the low-temperature region where the electron can propagate coherently with magnon. Moreover, we should note that the scaling between $\Delta Q$ and $\Delta S$ is worse above $T_c$. As a possible origin of this, we point out that the thermopower becomes almost independent of temperature towards 100 K, where the Heikes formula[4] seems to be better than the Kelvin formula, and $\Delta Q$ cannot be compared with $\Delta S$. Klein et al. have suggested that the temperature-independent thermopower at high temperatures can be explained by the Heikes formula in the related ruthenium oxide $SrRuO_3$.[6]

Finally, we will point out some significant implications from Fig. 5. (i) In conventional metals, the thermopower has been compared with the electron specific heat.[32] This is not always valid when the system undergoes a phase transition. The thermopower change should be compared with the entropy change near the transition temperature. (ii) In magnetic materials, the resistivity and the thermopower give complementary information, as was reported in $Sr_3YCo_4O_{10.5}$.[9] While the resistivity is susceptible to the change in scattering time, the thermopower detects the entropy change. (iii) Thus far, the Kelvin formula has been examined by comparing measured data with ab initio calculations, in which electronic-band effects and strong-correlation effects are inseparably involved.[15] It thus remains unexamined to what extent the Kelvin formula explains the strongly correlated part of the thermopower. In the present study, we have experimentally demonstrated that the magnetic entropy certainly contributes to the thermopower in accordance with Eq. (4). (iv) We find some similarity between $Sr_3YCo_4O_{10.5}$ and $CaRu_{0.8}Sc_{0.2}O_3$ in the sense that the thermopower detects the spin entropy in the background. Thus, as a microscopic picture of $CaRu_{0.8}Sc_{0.2}O_3$, we suggest that the $Ru^{5+}$ ions are randomly dispersed in ferromagnetic grains, and interact with the itinerant electrons in the $Ru^{4+}$ ions.

## 4. Summary

We have investigated the magneto-transport properties and the magnetocaloric effect in $CaRu_{0.8}Sc_{0.2}O_3$. A negative magnetoresistance is observed, which is ascribed to the tunnel magnetoresistance in which the carriers undergo spin-dependent scattering at the boundaries between the ferromagnetic domains. Unlike the resistivity, the thermopower is significantly suppressed around $T_c$ of 30 K by a magnetic field, clearly showing that the ferromagnetism affects the thermopower. We find that the magneto-entropy remarkably resembles the magneto-thermopower, and the relationship between the two can be analyzed using the Kelvin formula. It has been never tested experimentally whether or not the Kelvin formula can describe the strongly correlated part in the thermopower. The present study has clarified that the Kelvin formula quantitatively explains the magnetic part of the thermopower in the weak ferromagnet $CaRu_{0.8}Sc_{0.2}O_3$. The technique proposed in this paper can be applied to other magnetic materials to examine the validity of the Kelvin formula.

**Acknowledgment**

This work was partially supported by Grants-in-Aid for Scientific Research, MEXT,Japan (Nos. 25610091 and 26247060). One of the authors (T. D. Y.) was supported by the Program for Leading Graduate Schools ``Integrative Graduate Education and Research in Green Natural Sciences", MEXT, Japan. A part of this work was performed under the Inter-University Cooperative Research Program of the Institute for Materials Research, Tohoku University (Proposal Nos. 15K0021 and 14K0024).

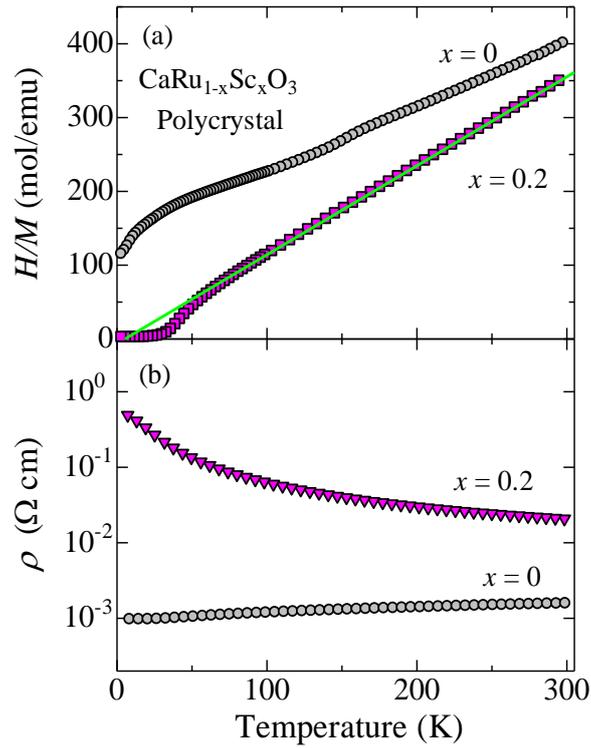

**Fig. 1.** Temperature dependence of (a) the reciprocal magnetic susceptibility $H/M$ and (b) the resistivity $\rho$ for CaRu$_{1-x}$Sc$_x$O$_3$ ($x = 0, 0.2$). The solid line depicts a Curie-Weiss fit for $H/M$ of the $x = 0.2$ sample.

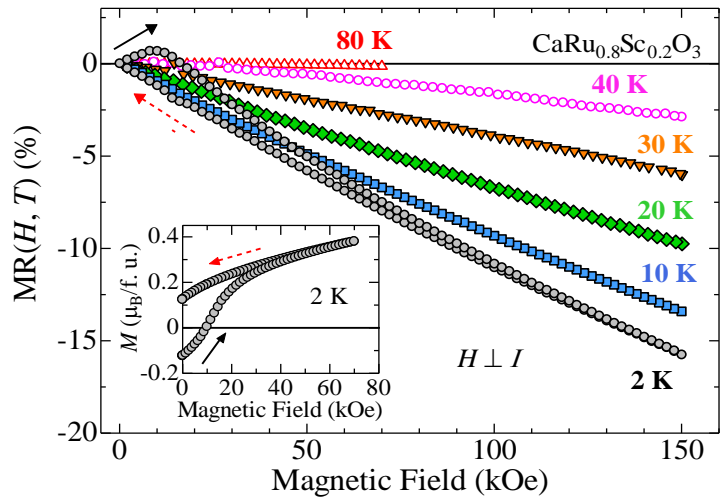

**Fig. 2.** Field dependence of the magnetoresistance MR($H$, $T$) at various temperatures. The inset shows the field dependence of the magnetization $M$ at 2 K in the field range of $0 \leq H \leq 70$ kOe. The solid and broken arrows represent the field-increasing and field-decreasing processes, respectively.

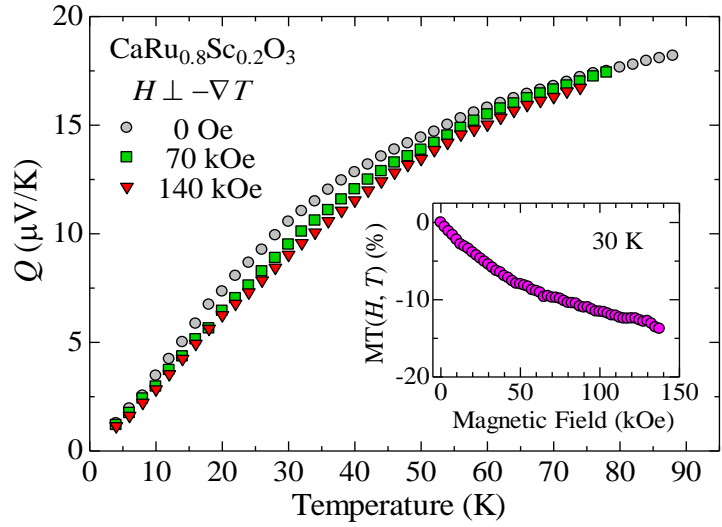

**Fig. 3.** Temperature dependence of the thermopower $Q(H, T)$ in several magnetic fields. The inset shows the field dependence of $MT(H, T) = Q(H, T)/Q(0, T) − 1$ at 30 K.

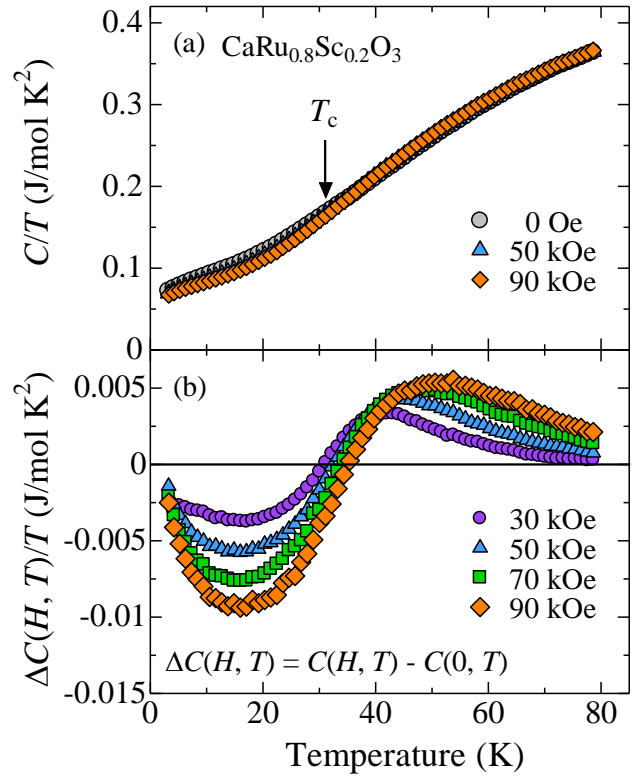

**Fig. 4.** Temperature dependence of (a) the specific heat divided by temperature, $C/T$, and (b) $\Delta C(H, T)/T$ for different magnetic fields.

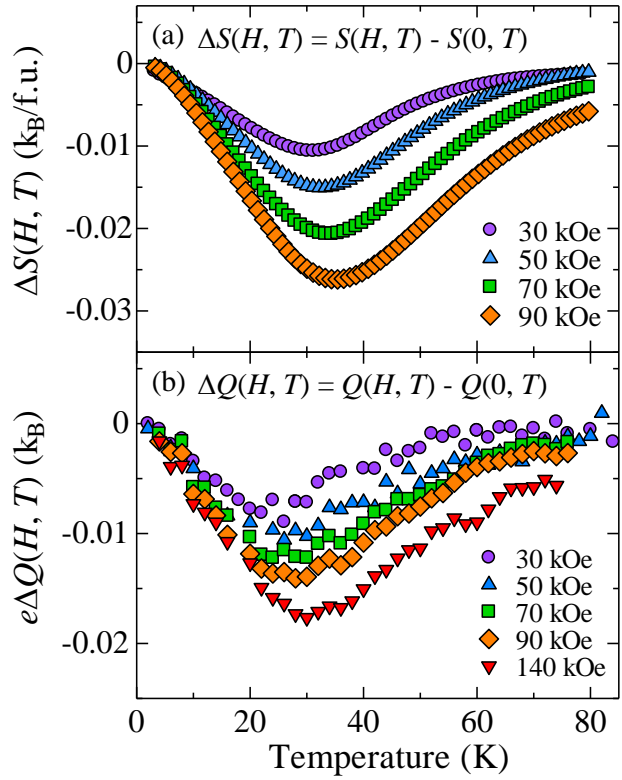

**Fig. 5.** (a) Magneto-entropy $\Delta S(H, T)$ plotted in units of $k_B$ per formula unit. The data are calculated from $\Delta C(H, T)/T$ (see text). (b) Magneto-thermopower plotted as $e\Delta C(H, T)$ in units of $k_B$.

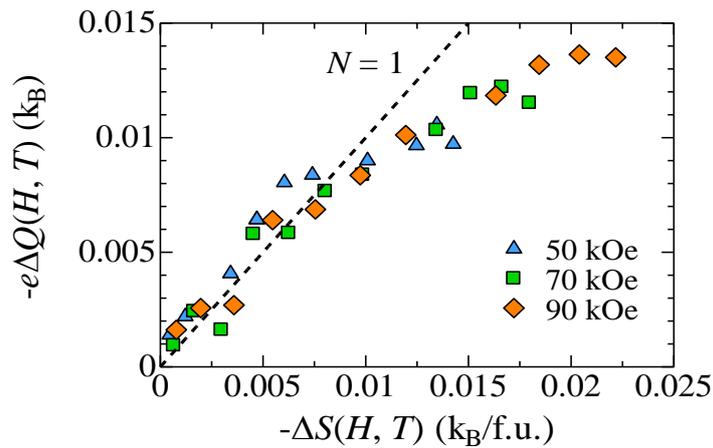

**Fig. 6.** Magneto-thermopower $\Delta Q(H, T)$ plotted as a function of magneto-entropy $\Delta S(H, T)$ below $T_c$. The dotted line represents $N = 1$ for the Kelvin formula.